\documentclass[%
 reprint,
 amsmath,amssymb,
 aps,
]{revtex4-2}

\usepackage{graphicx}
\usepackage{dcolumn}
\usepackage{bm}
\usepackage{appendix}
\usepackage{float}
\usepackage{subfigure}

\begin{document}

\preprint{APS/123-QED}

\title{Sublattice Melting in Binary Superionic Colloidal Crystals}

\author{Yange Lin}
\affiliation{
 Department of Chemistry, Northwestern University, Evanston, IL,
60208
}
\author{Monica Olvera de la Cruz}%
 \email{m-olvera@northwestern.edu}
\affiliation{
 Department of Chemistry, Northwestern University, Evanston, IL, 60208\\Department of Physics and Astronomy, Northwestern University, Evanston, IL, 60208\\Department of Materials Science and Engineering, Northwestern University, Evanston, IL, 60208
}

\date{\today}

\begin{abstract}
In superionic compounds one component pre-melts providing high ionic conductivity to solid state electrolytes. Here, we find sublattice melting in colloidal crystals of oppositely charged particles that are highly asymmetric in size and charge in salt solutions. The small particles in ionic compounds melt when the temperature increases forming a superionic phase. These delocalized small particles in a crystal of large oppositely charged particles, in contrast to superionic phases in atomic systems, form crystals with non-electroneutral stoichiometric ratios. This generates structures with multiple domains of ionic crystals in percolated superionic phases with adjustable stoichiometries.
\end{abstract}

\maketitle

Colloids of various components have been assembled into diverse crystalline structures \cite{Pine,Talapin,Tezcan}, and have served as experimental models to study phase behaviors \cite{phasetransition2,phasetransition3} and self-assembly processes \cite{Pine,Chaikin}. Unlike chemical compounds in atomic systems, colloidal assemblies do not have constraints from the number, the symmetry, or the energy of orbitals. This significantly diversifies possible crystal structures. In the past few decades, several types of binary colloidal crystals with different component ratios have been studied, such as $AB$ \cite{AB}, $AB_2$ \cite{AB2}, and $AB_8$ \cite{AB8}. Experiments and computer simulations have shown that the size ratio \cite{sizeratio&entropydriven} or charge ratio \cite{AB8, chargeratio} of the two components as well as the ionic strength in the solution \cite{AB8} are important factors in the assembling process of binary superlattices. Most of the work on binary charged colloidal crystals retain the restricted structures of classical atomic ionic compounds in which all particle positions are fixed. 

In contrast, atomic ``superionics" such as superionic conductors possess different kinds of structures and properties. In superionic conductors, due to the low energy barrier along cation migration paths \cite{superionicconductor1}, one of the components, termed the fast ions, are mobile and have a delocalized density distribution within the crystal lattice \cite{superionicconductor1,superionicconductor2}. As a result, superionic conductors have high ionic conductivity at room temperature, making them the core component of high-performance solid state batteries. Superionic phases also have been found in other microscopic condensed systems such as ammonia \cite{ammonia}, ice \cite{ammonia,ice}, and polymers \cite{polymer}. 

Here, we explore the possibility of assembling superionic conductors from oppositely charged colloids in salt solutions. The charge neutrality restriction in traditional ionic and superionic atomic crystals can be removed in charged colloidal crystals when the screening from small ions is present \cite{AB8,chargedcolloid1}. Moreover, the range of the interaction potential can be readily tuned by controlling the salt concentration in regimes where the Debye-Huckel approximation is valid \cite{chargedcolloid1,chargeratio}. Therefore, these colloidal systems may substantially expand the scope of colloidal science and superionic materials. Recently, a related superionic-like phase was found in binary colloidal mixtures of large DNA-functionalized gold nanoparticles (DNA-AuNPs) and complementary small DNA-AuNPs grafted with free strands that hybridize only to the large DNA-AuNPs \cite{metallicity}. While the simulation predictions were for monodispersed samples at zero external pressure \cite{MartinPhD_Ch7_128}, in the experiments highly polydispersed small DNA-AuNPs were used, and free linkers that could act as depletants were added \cite{metallicity}. Therefore, due to the experimental limitations we cannot neglect the depletion effects in these systems. Furthermore, the nature of the transition with the temperature was not determined in these studies. Instead, by considering monodispersed colloidal charged particles without grafted linkers the possibility of transitions from ionic to  superionic phases can be analyzed.  In this paper, we find a sharp transition from ionic to superionic phases in charged colloidal crystals characterized by a discontinuous jump in the lattice spacing as the temperature increases, as well as by the double-well shape of the free energy landscape via molecular dynamics (MD) simulations. Moreover, we find regions of coexistence between phases such as ionic-like phases of different stoichiometries at low temperatures, or ionic-like phases coexisting with superionic-like phases at intermediate temperatures, which we analyze by calculating the time average density of the small particles \cite{OlveraPRL1991}. Finally, we show that the attractions provided by the small particles in superionic-like phases are not depletion type interactions.

Since colloidal mixtures of oppositely charged components with similar sizes form ionic phases \cite{AB8, chargedcolloid1,chargeratio}, we concentrate our study in binary oppositely charged colloids whose components differ in charge and size substantially. In our MD simulations, both large (A) and small (B) particles are modeled as isotropic, charged spheres. The interactions between same species are repulsive thus they cannot form crystal structures from pure As or Bs in the absence of external pressure which is the case simulated here. All the ions are accounted for implicitly by applying the Debye-Huckel approximation, which describes pair potentials between charged nanoparticle at salt concentrations up to roughly 300mM of NaCl \cite{depletion3}. Particles interact through the Weeks-Chandler-Andersen (WCA) potential for excluded volume effects and the Debey-Huckel (DH) potential for the screened Coulombic interactions:

\begin{equation}
U(r)=U_{WCA}(r)+U_{DH}(r)
\end{equation}
\begin{equation}
  U_{WCA}(r_{ij})=\begin{cases}
    4\varepsilon\left[\Big(\dfrac{\sigma_{ij}}{r_{ij}}\Big)^{12}-\Big(\dfrac{\sigma_{ij}}{r_{ij}}\Big)^{6}\right]+\varepsilon, & r_{ij}<r_{cut}.\\
    0, & r_{ij}>r_{cut}.
  \end{cases}
\end{equation}
\begin{equation}
 U_{DH}(r_{ij})=\frac{q_i^*q_j^*e^{-\kappa r_{ij}}}{r_{ij}/\sigma}\varepsilon
\end{equation}
The energy term of the WCA potential, $\varepsilon$, is chosen to be the characteristic energy parameter in our simulations. The cutoff distance of the WCA potential $r_{cutWCA}^{ij}=2^{1/6}\sigma_{ij}$, where $\sigma_{ij}$ is pair-dependent and is calculated from the Lorentz-Berthelot mixing rules $\sigma_{ij}=R_i+R_j$. Here the radii of the two species were fixed at $R_A=5\sigma$ and $R_B=1\sigma$, where $\sigma$ is the characteristic distance parameter. For the Debey-Huckel potential, $\kappa$ is the screening strength and $q_i^*$ and $q_j^*$ are effective reduced charges. For colloidal particles, an extended form of $q_i^*$ commonly used in simulations includes the hard-core via the DLVO potential which gives $q_i^*=q_ie^{\kappa R_i}/(1+\kappa R_i)$ \cite{AB8,YukawaFromsCrystals-bier2010phase}, yet it is accurate only for dilute systems. In concentrated colloidal suspensions, such as in the crystals studied here, $q^*$ has a more complicated form \cite{EffectiveCharge}. Thus, without losing generality we directly use $q^*$ as simulation parameters that are independent of $\kappa$ and the compactness of the system. We keep the exponentially-decaying part with the distance between particles, because it is preserved in nonlinear models even when water effects and ions are explicitly included \cite{depletion3}. The cutoff distance of the Debey-Huckel potential $r_{cutDH}^{ij}=3\kappa^{-1}+\sigma_{ij}$.

From the energy unit $\varepsilon$ and distance unit $\sigma$, the reduced quantities can be defined, including the reduced temperature $T^*=kT/\varepsilon$, reduced pressure $P^*=P\sigma^3/\varepsilon$, reduced time $\tau^*=t\sqrt{\varepsilon/(m\sigma^2)}$, and reduced charges $q^*=q/\sqrt{4\pi\epsilon_0\epsilon_r\sigma\varepsilon}$ where $e$ is the elementary charge and $\epsilon_0$ and $\epsilon_r$ are the dielectric constants of the vacuum and the media, respectively. In the rest of the paper the prefix "reduced" will be omitted and these quantities are in terms of the reduced quantities.

\begin{figure}[b]
\includegraphics[width=\linewidth]{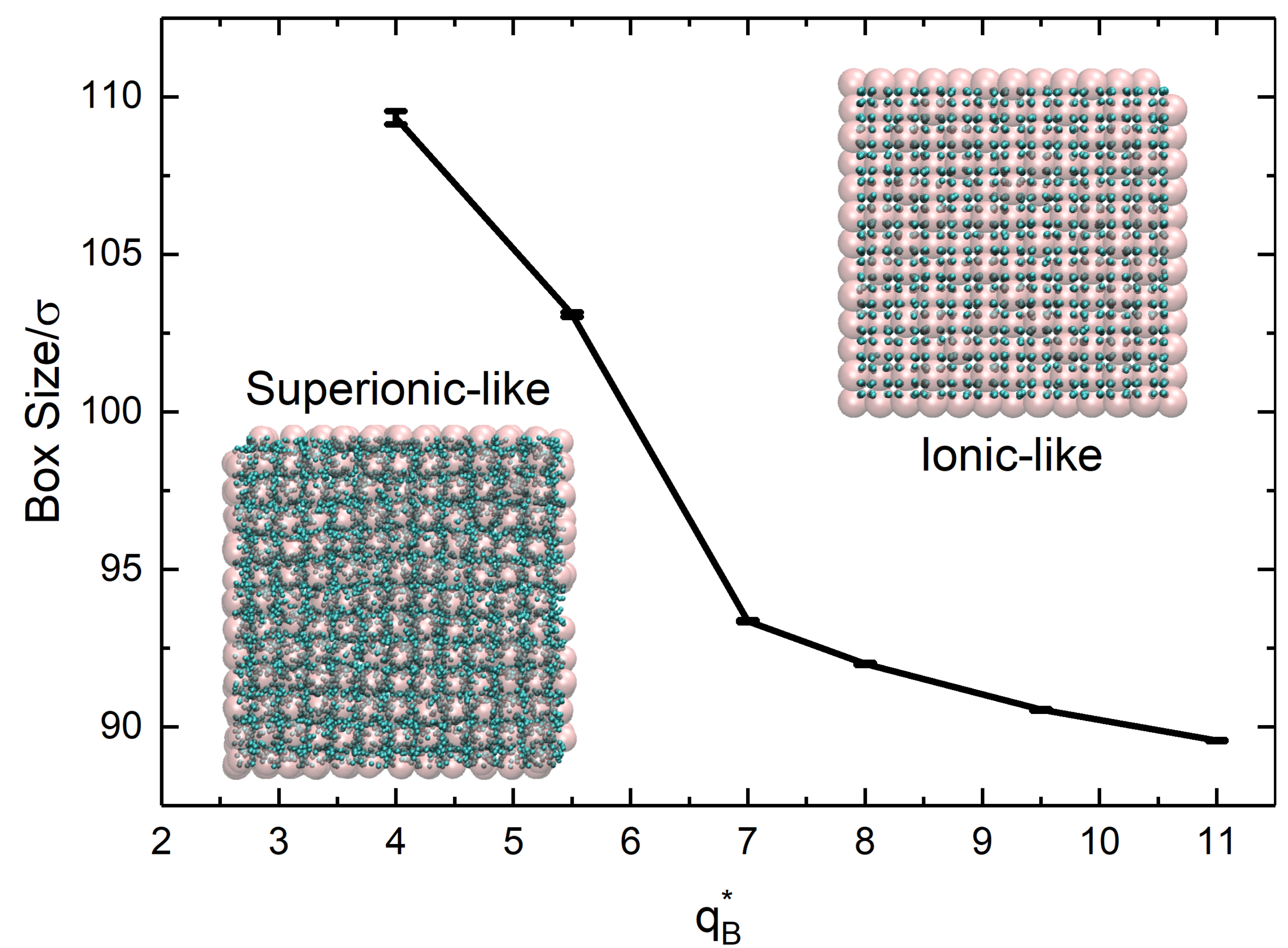}
\caption{\label{fig:epsart} Simulation box size under different reduced charges of the small particles ($q_B^*$) at $T^*=0.3$, for $N_B/N_A=8$, $q_A^*=-247$ at $\kappa\sigma=0.7$. Two distinct phases are observed, superionic-like and ionic-like; the snapshots are from the [001] direction of the FCC crystal.}
\end{figure}

All the molecular dynamic (MD) simulations are done in the LAMMPS software package \cite{LAMMPS}. In the zero pressure NPT simulation, we initialize the system by setting large particles in perfect face centered cubic (FCC) crystal positions in a periodic cubic box, with small particles randomly placed throughout the lattice while avoiding strong overlap. The number of crystal unit cells in each direction is $6$ (We have examined larger systems and found that the finite size effect is negligible. See Fig. 5 in the Appendix for more details). The system is first thermalized in the canonical (NVT) ensemble with a Langevin thermostat, then is slowly compressed to a close-packing state by reducing the simulation box size. After the system is equilibrated for $2\times10^3 \tau^*$ ($10^6$ timesteps), it is switched to the isobaric\textendash isothermal (NPT) ensemble with a large enough external pressure to keep the system compressed and run for another $2\times10^3 \tau^*$. The pressure is subsequently relaxed to exactly 0 and the system is further equilibrated for $2\times10^4 \tau^*$ ($10^7$ timesteps). To simplify the simulations, the cubic symmetry of the simulation box is maintained during the run. Removal of this symmetry constraint may allow the crystal to transfer from FCC to other non-cubic structures, or to other cubic structures such as body centered cubic (BCC) more easily, however, these additional complexities are beyond the scope of the current paper. 

Our results show that under mediate salt conditions, by reducing the attraction strength between the two components (A-B attraction) the colloidal crystals can transit from ionic phases to superionic phases. In Fig. 1 we explore how the equilibrium size of the simulation box varies with the reduced charge of small particles, $q_B^*$. Here $T^*=0.3$, $N_B/N_A=8$, $q_A^*=-247$, and $\kappa\sigma=0.7$ \footnote{If we choose $\sigma=1nm$, this screening strength would approximately correspond to a salt concentration of $44mM$ NaCl, which is within the concentration range where the Debye-Huckel approximation is applicable. Moreover, the large and the small nanoparticle sizes would be 10nm and 2nm, respectively, which are also reasonable values in experiments.} (Hereafter, the stoichiometric ratio of small (B) and large (A) particles $N_B:N_A$ is given by $N_B/N_A$). When $q_B^*=+11$ the particles aggregate into an ionic crystal in which small particles are fixed at interstitial positions and form a regular sublattice. As $q_B^*$ decreases, the A-B attraction decreases and the equilibrium box size gradually increases. When $q_B^*=+5.5$ and $+4$ the attraction strength is no longer sufficient to localize the small particles at specific positions but is still able to keep the crystal stable. Therefore the sublattice melts and the system transits to a superionic-like structure. Further decreasing the attractions by using either smaller $q_B^*$ or larger $\kappa$ induces the melting of the whole FCC crystal. Moreover, increasing the A-B attraction by reducing the salt concentration ($\kappa\sigma=0.1$) also leads to the crystal melting because the repulsion between large particles is enhanced and dominates. This results in an equilibrium gas state where large particles stay far apart with small particles surrounding each of them. An example of how an unstable crystal melts as the pressure approaches $0$ is shown in Fig. 6 in the Appendix.

\begin{figure}[b]
\includegraphics[width=\linewidth]{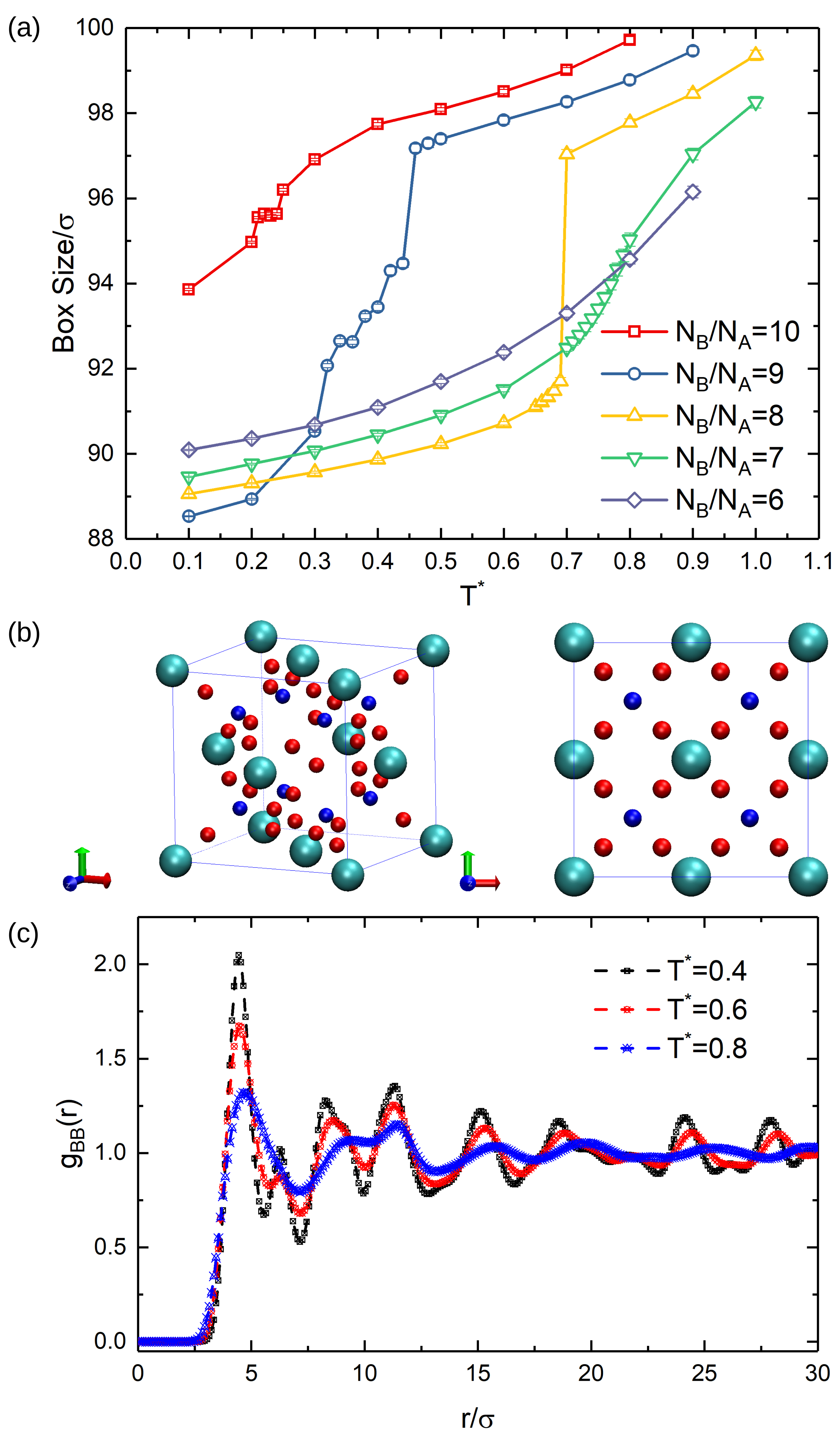}
\caption{\label{fig:epsart} (a) Heating curves of the box size under different number ratios $N_B/N_A$ with $q_A^*=-247$, $q_B^*=+11$, $\kappa\sigma=0.7$. (b) Schematic plots of the $32f$ and $8c$ Wyckoff positions in the FCC crystal. (Left) One FCC unit cell contains four atoms (cyan), eight $8c$ positions (blue), and thirty\textendash two $32f$ positions (red). (Right) The view from the [001] direction. (c) The radial distribution function of small particles $g_{BB}(r)$ at $N_B/N_A=7$ for $T^*=0.4$, 0.6, and 0.8.}
\end{figure}

By increasing the temperature, $T^*$, above $0.3$ in the system with $q_A^*=-247$, $q_B^*=+11$, and $\kappa\sigma=0.7$, we also observe sublattice melting and that this ``ionic-superionic" transition is strongly first order at $N_B/N_A=8$. Note that if we use convert $q_i^*$ to bare charges $q_i$ using DLVO approximation for the above parameters, we obtain $q_A\approx-34$ and $q_B\approx+9$; as a reference, when $q_B^*=+5.5$ the crystal is nearly electroneutral ($q_A\approx-34$ and $q_B\approx+4$), and we also observe sublattice melting shown in Fig. 1. Comparing these two crystals demonstrates that electroneutrality is not a requirement for sublattice melting providing there is enough screening. To determine the sublattice melting temperature, we analyze changes in the equilibrated simulation box size which is approximately six times the lattice spacing. Heating curves of the box size with different number ratios $N_B/N_A$ (Fig. 2a) show that the lattice expands as the temperature increases, however, at $N_B/N_A=8$ the expansion is discontinuous at a certain temperature ($T^*=0.68$). The discontinuous jump in the magnitude of the lattice spacing, which corresponds to sublattice melting of small particles, indicates that this melting occurs via a first order phase transition. A similar but weaker discontinuous lattice expansion occurred at $N_B/N_A=10$. For $N_B/N_A=9$ the lattice has two distinct discontinuous expansions (at $T^*=0.3$ and $T^*=0.46$), which later we find that are caused by two separate sublattice melting in two ionic phases with different favorable stoichiometric ratios ($N_B/N_A=8$ and $N_B/N_A=10$).

We can use a fundamental concept in crystallography, Wyckoff positions, to understand why this first order phase transition happens only at $N_B/N_A=8$ and $N_B/N_A=10$. Wyckoff positions are widely used for the determination and description of crystal structures. They describe the positions of special sites and their symmetries inside a unit cell \cite{WyckoffPosition}. Here we use them to describe where the small particles can be found in an size-asymmetric binary crystal. In the FCC unit cell, there are two important Wyckoff positions, the $32f$ and the $8c$ positions (Fig. 2b). As their names stated, one FCC unit cell contains 8 $c-$positions ($C-$centered positions) and 32 $f-$positions (face centered positions). Since one FCC unit cell also contains 4 large particles, the number ratio between the $32f$ positions and the large particles is $8/1$, and between the $8c$ positions and the large particles is $2/1$. Note that there are 8 tetrahedral voids inside the FCC unit cell, and the $8c$ positions are the centers of these tetrahedrons, while the $32f$ positions are the four inner face centers of these tetrahedrons (such that there are $4 \times 8=32$ of them). In the size-asymmetric oppositely-charged binary crystal, the small and lower in charge particles on the $32f$ positions have lower energy than on the $8c$ positions, since each particle on the $32f$ positions is closer to oppositely-charged large particles than the $8c$ positions. Therefore, in the ionic phases, the small particles tend to first occupy the $32f$ positions and then the $8c$ positions, and fulfilling them will result in two favorable number ratios $N_B/N_A=8$ and $N_B/N_A=10$ (In fact, the $32f$ positions are commonly the home for the small particles in a $AB_8$ binary ionic crystal \cite{AB8, chargeratio, chargedcolloid1}). At these number ratios and at low temperatures, the crystal is in the ionic state that is enthalpically-favorable but entropically-unfavorable because the crystal has little defects and the small particles are basically stuck at their equilibrium places, which is a great entropy lost comparing to the superionic state in which the small particles can access to the whole free space inside the crystal. As a result, there exists a transition temperature above which the system favors entropy over enthalpy and expands the lattice spacing for small particles previously trapped in the interstitial positions to delocalize (see Fig. 7 in the Appendix for how the diffusion coefficient of small particles varies with the temperature).

Based on the volume expansion, the ``ionic-superionic" transition seems continuous at other number ratios $N_B/N_A<8$. A possible explanation for it is that the volume expansion is mainly governed by the thermal expansion in those systems, because there are vacancy defects (unoccupied $32f$ positions) and the cohesive energy is lower. These superstructures, when in their ionic state, resemble the interstitial solid solution (ISS) phase found in size asymmetric hard sphere mixture under large external pressure \cite{laura}. In the ISS phase, the large spheres form the crystal lattice, and the small spheres, whose number is less than the number of interstitial sites, partly occupy the interstitial sites and can diffuse among these sites through vacancies, akin to a fluid. To evaluate the nature of the transition, an order parameter, generally the density fluctuation around the mean density $\delta \rho$, and the correlation length analyses are required. Change in symmetry of this order parameter in systems transitioning from isotropic (or delocalized in space) to periodic (localized on lattice sites) structures by decreasing the temperature cannot be continuous \cite{brazovskii}. In Fig. 2c we examine the radial distribution function of small particles $g_{BB}(r)$ in the crystal with $N_B/N_A=7$ at different $T$ values, and show that there is a symmetry change since the long range ordering and the peak positions are different in the superionic ($T^*=0.8$) and ionic phases ($T^*=0.4$). Therefore, it is possible that sublattice melting at ratios $N_B/N_A < 8$ is weakly first order.

\begin{figure}[b]
\includegraphics[width=\linewidth]{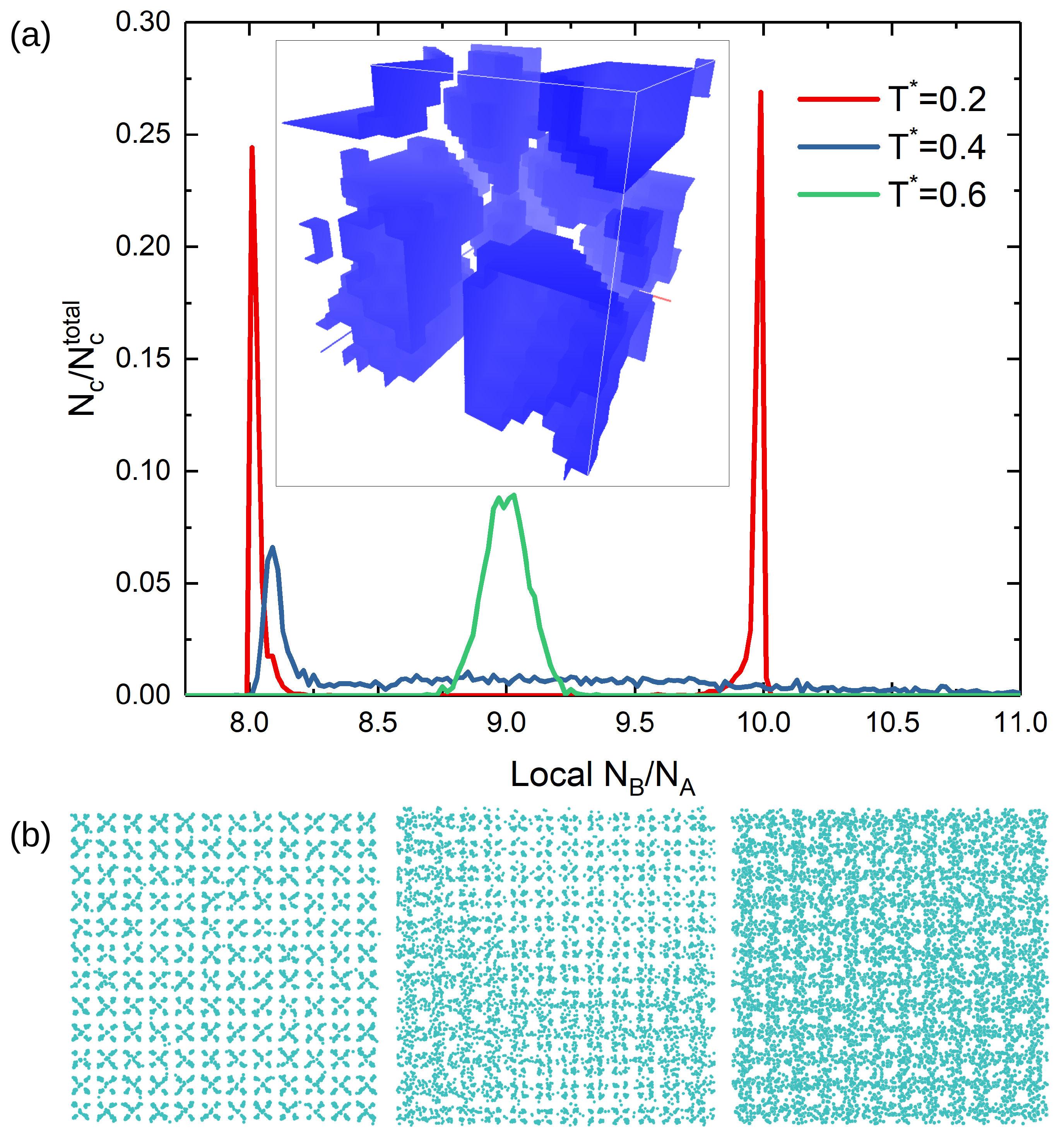}
\caption{\label{fig:epsart} Phase coexistence in the system with $N_B/N_A=9$. To obtain better statistics, we enlarged the simulation box to $8\times8\times8$ unit cells. (a) The histogram of average number of small particles in sub-unit cubic bins at different temperatures $T^*=0.2$, 0.4, and 0.6. $N_C$ is the number of cubes that has a certain local $N_B/N_A$ and $N_c^{total}$ is the total number of cubes which is $16^3=4096$ here. (Inset) A 3-d view of the simulation box showing locations of cubes with local $N_B/N_A=8$ inside the crystal at $T^*=0.4$. All the cubes satisfying $\left|N_B/N_A-8\right|<0.2$ are colored blue while the rest are left blank. (b) Snapshots of the equilibrium distribution of small particles at $T^*=0.2$ (left), $0.4$ (middle), and $0.6$ (right).}
\end{figure}

For $N_B/N_A=9$ systems, at low temperatures, we observe two coexisting ionic phases with local number ratios $N_B/N_A=8$ and $N_B/N_A=10$, and each of these phases transitions into superionic phases at different temperatures as the temperature increases. To obtain the local number ratios, we divided the simulation box into small cubic bins, which have 1/8 the volume of the FCC unit cell and are the smallest chemically identical unit for small particles. After equilibrium, the average number of small particles in each cube $N_B^{local}$ was counted from 1000 frames taken every 5000 timesteps ($10\tau^*$) and local number ratio is then given by $N_B/N_A=2N_B^{local}$ as one cube has 1/2 large particle. The histogram of local $N_B/N_A$ at different temperatures $T^*=0.2$, 0.4, and 0.6 combined with corresponding simulation snapshots (Fig. 3) reveal that at $T^*=0.2$ the system consists of two kinds of ionic crystals with stoichiometric ratios $N_B/N_A=8$ and $N_B/N_A=10$, respectively. These two ionic phases are both in micro-size (Fig. 8 in the Appendix). However, because the $8c$ positions have higher energy than the $32f$ positions, the $N_B/N_A=10$ ionic phase has a lower sublattice melting temperature than the $N_B/N_A=8$ phase. Therefore, when the temperature is raised to $0.4$, the $N_B/N_A=10$ phase melts into the superionic state and we observe the $N_B/N_A=8$ ionic phases coexisting with superionic phases that have various local number ratios distributed almost evenly in a wide range. By plotting the locations of cubes with local $N_B/N_A=8$ we find that instead of aggregating into a macro\textendash crystal, these ionic cubes form microphases scattered throughout a percolated structure of superionic phases (the cluster sizes span from 2 to 6 unit cells in our simulations) probably to decrease the surface strain generated from the lattice constant mismatch between the ionic and the superionic phases (see Fig. 2a). Further increasing the temperature melts the sublattice in the $N_B/N_A=8$ phase and the whole system forms a homogeneous superionic phase with $N_B/N_A=9$. From the phase coexistence information, it is clear that the most stable stoichiometry for the ionic phase in FCC crystals is $N_B/N_A=8$, but we do not know if there is an optimal stoichiometric ratio for the superionic phase since that would require equal chemical potential simulations. It is important to note that with various possible stoichiometric ratios the system may end up into a glass state in which the large particles are fixed while the small ones are fluid, similar to what has been predicted in the charge- and size-asymmetric ionic system with coulombic interactions \cite{glass}. In the current work, we have restricted the crystal to the cubic symmetry with a fixed stoichiometry. Therefore, the equilibrium structure may not represent the most stable state when the box symmetry restriction and stoichiometric constraint are removed, such as in the case of deformable crystals that can exchange components with the surroundings.

Relative Helmholtz free energy landscapes are calculated by thermodynamic integration methods \cite{frenkel}. In thermodynamics, the Helmholtz free energy, $F$, is related to the pressure by $-P=\left(\partial F/\partial V\right)_{N,T}$. Therefore, the relative free energy can be calculated from the integral:

\begin{equation}
\begin{split}
F_{rel}(V) & =F(V)-F(V_0)=-\int_{V_0}^V PdV \\
& \approx -\sum_i(P_{i+1}+P_i)(V_{i+1}-V_i)/2
\end{split}
\end{equation}
where $F(V_0)$ is the reference state, and midpoint approximation was used to numerically evaluate the integral.

\begin{figure}[t!]
\includegraphics[width=\linewidth]{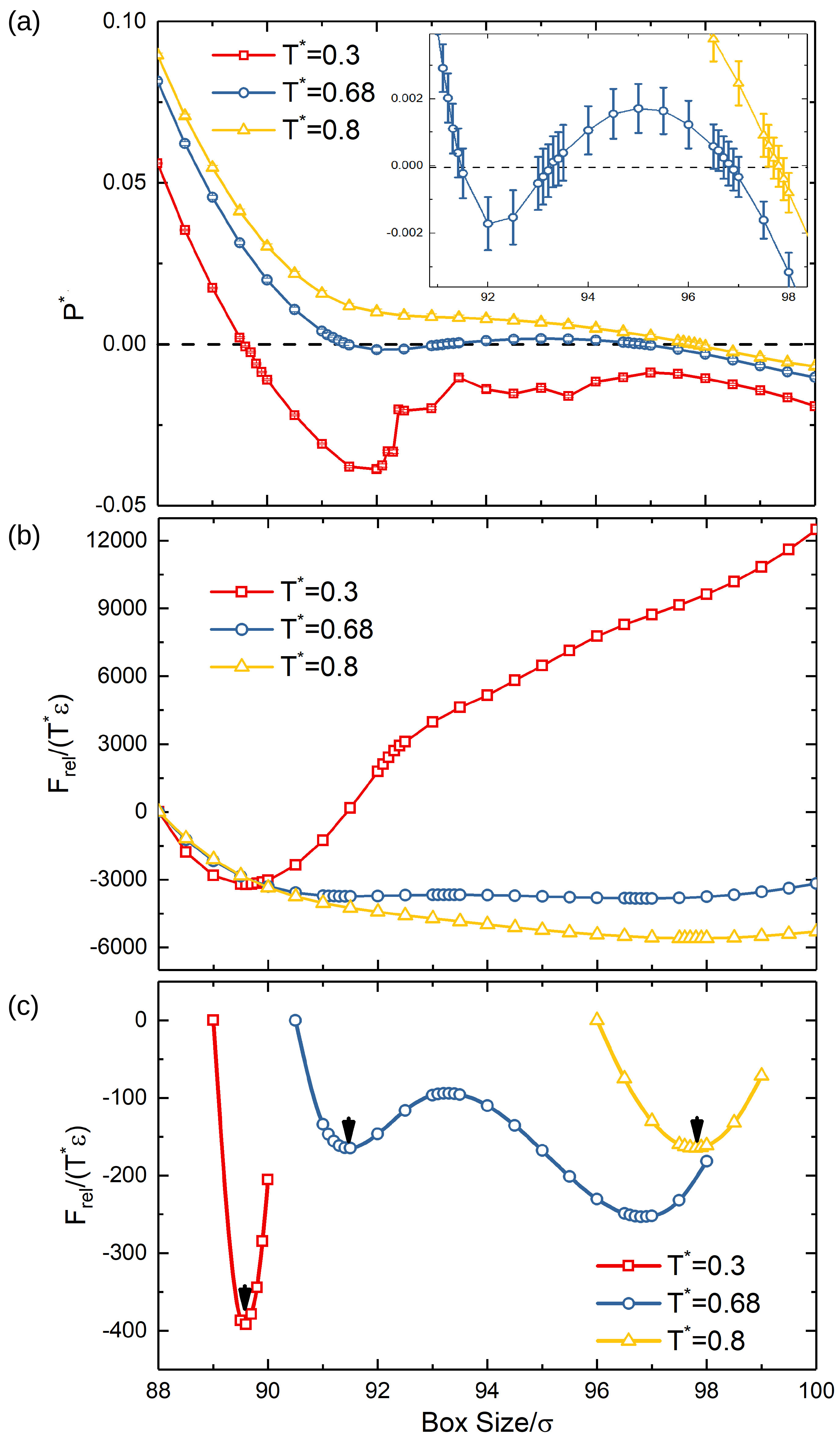}
\caption{\label{fig:epsart}  Calculations of the relative Helmholtz free energy in the $N_B/N_A=8$ systems at different temperatures ($T^*=0.3$, $0.68$, and $0.8$) via thermodynamic integration methods. (a) The Pressure\textendash Volume plot obtained in the NVT simulations. The enlarged view of the zero points of the pressure are shown in the Inset. (b) Overall landscape of relative Helmholtz free energy. The reference volume $V_0$ is $(88\sigma)^3$ for all three temperatures. (c) Locations of free energy minimums. The reference volumes are $(89\sigma)^3$ for $T^*=0.3$, $(90.5\sigma)^3$ for $T^*=0.68$, and $(96\sigma)^3$ for $T^*=0.8$. Black arrows mark the state sampled by previous NPT simulations.}
\end{figure}

A series of NVT simulations with a Langevin thermostat are done for different volumes ${V_i}$ at $N_B/N_A=8$. The system is first initialized in the same way as in the NPT simulations. After thermalized in the NVT ensemble with a Langevin thermostat for $2\times10^3 \tau^*$, the initially large simulation box is deformed to the volume $V_i$ and further equilibrated for $2\times10^4 \tau^*$ to obtain the corresponding ensemble averages of pressure $P^*_i$ (Fig. 4a). The curve at $T^*=0.68$ in Fig. 4a resembles the van der Waals loop. However, this curve results from the finite\textendash size of simulation box (which means the loop on this curve will reduce to a flat line in an infinite system at the equilibrium) \cite{loop}. Negative pressures in the simulations mean that the system tends to aggregate.

Plugging the pressure and volume data in Eq.4 we obtain the relative free energy landscapes for the system with $N_B/N_A=8$ at different temperatures (Fig. 4b and 4c). The curves are plotted in $F_{rel}/(T^*\varepsilon)=F_{rel}/kT$ in order to better compare the depth of minimums with the termal motion. In the thermodynamic integration, the points where the pressure goes to zero correspond to extrema in the free energy landscape. Generally, one zero in the pressure corresponds to one well in the landscape, and three zeros correspond to two wells and one maximum in the landscape. At both low $(T^*=0.3)$ and high $(T^*=0.8)$ temperatures, the free energy has only one minimum in the compact state, marking the ionic and superionic phases, respectively. The double-well shape around the transition temperature $(T^*=0.68)$ confirms that sublattice melting is a first order phase transition when the system is at the optimal stoichiometry. Volumes at the free energy minimums match well with the equilibrium volumes obtained in previous NPT simulations, although when there are double wells, NPT simulations tend to sample the state with smaller volume because we initialized the system in denser configurations.

Depletion forces are widely recognized to drive the assembly of mixtures of colloidal particles with different sizes \cite{sizeratio&entropydriven,depletion4-dijkstra1999direct}, but they are not important in stabilizing the superionic structures found here. First, in our simulations we do not have explicit salt which can cause depletion attraction between the nanoparticles \cite{depletion3}, and, in relation to experiments, provided the experiments are done at 300mM of NaCl or less there is no evidence of monovalent salt mediated attractions (even in large colloids provided the colloids have sufficient charge \cite{Zwanikken}). Second, depletion is mainly entropy-driven and should be enhanced by increasing temperature. However in our simulations, all colloidal crystals melt into gas phases when the temperature is increased above 1.3. Third, because the box size is not constrained in our zero pressure NPT simulation, the system is supposed to expand infinitely if it was favorable to add more free volume for the small particles, but instead the system is equilibrated at a finite size. The average distance between two neighboring large particles $d$ in our simulations satisfies $2R_A<d<2\sigma_{AB}$ where $\sigma_{AB}=R_A+R_B$. Depletion effects can exist when $d$ is in the interval $\big(2R_A,2\sigma_{AB}\big)$. However, the free volume for small particles $V_{free}$ as a function of $d$ in the FCC structure is given by $V_{free}(d)\propto d^3-2\pi (R_A+R_B)^3/3+\pi(R_A+R_B-d/2)^2\big(4(R_A+R_B)+d\big)$ which monotonically increases in the interval $\big(2R_A,2(R_A+R_B)\big)$. Thus the colloidal superionic structure is not stabilized at any local maximum of $V_{free}$. 

To conclude, we have identified a superionic-like crystal structure in size-asymmetric charged colloidal systems where the smaller particles melt and hold the larger particles in a crystalline lattice via screened Coulomb interactions. By cooling down the system, the small mobile particles condense to interstitial positions, resulting in an ionic-like structure. At the stoichiometric ratio where the number of small colloids equals the number of interstitial positions, this colloidal``superionic-ionic" transition is first order, demonstrated by the discontinuous change in lattice constant and the double-well shape in the free energy landscape. The addition of more small colloids inside the lattice leads to the coexistence of ``ionic-like" domains and percolated``superionic-like" phases with multiple stoichiometries. This state of the system may provide insights for growing heterostructures. Overall, our findings provide guidelines to assemble metallic or superionic conductor colloidal crystals and set up the foundation for discovering exciting properties and functions of multicomponent colloidal crystals.

{\it Acknowledgement:}
This work has been funded by NSF DMR Award No. $1611076$. We thank Wei Li, Martin Girard, and Trung Nguyen for helpful discussions. We also thank the computational support of Sherman Fairchild Foundation.


\section*{Appendix}
\subsection*{1. Finite Size Effects}
\begin{figure}[H]
\centering
\includegraphics[width=0.95\linewidth]{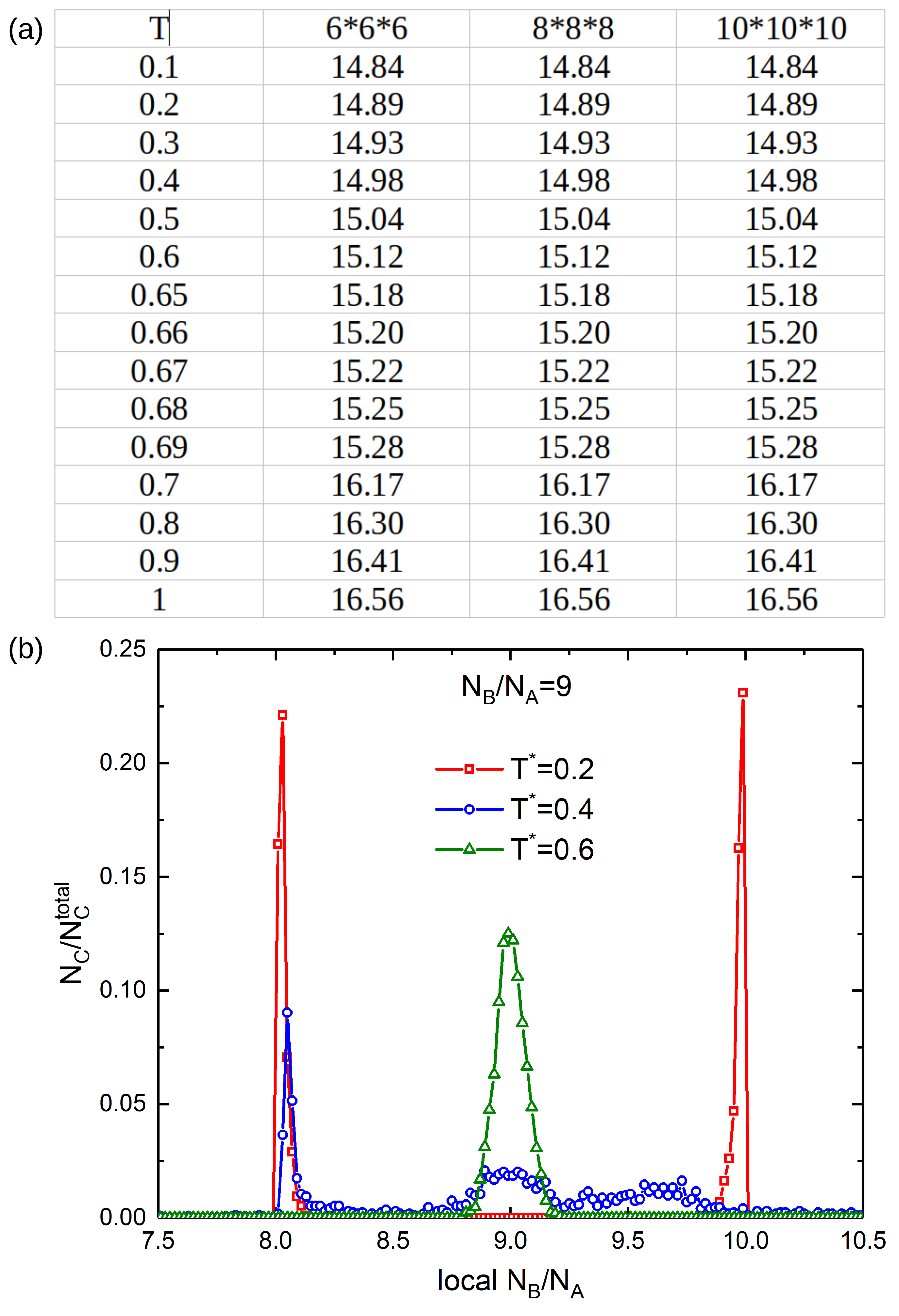}
\caption{Tests of finite size effects. (a) The equilibrium lattice constants (in $\sigma$) obtained in the zero pressure NPT simulations under different temperatures at $N_B/N_A=8$ using simulation boxes with the sizes $6\times6\times6$, $8\times8\times8$, and $10\times10\times10$ FCC unit cells, respectively. The results are exactly the same regardless of the crystal size. (b) The histogram of average number of small particles in sub-unit cubic bins at $T^*=0.2$, $0.4$, and $0.6$ using a $6\times6\times6$ unit cell simulation box, which is similar to the results obtained from an $8\times8\times8$ unit cell simulation box in Fig. 3. Therefore there is no significant finite size effects in our results.}
\end{figure}

\subsection*{2. The Crystal Melting}
\begin{figure}[H]
\centering
\includegraphics[width=0.95\linewidth]{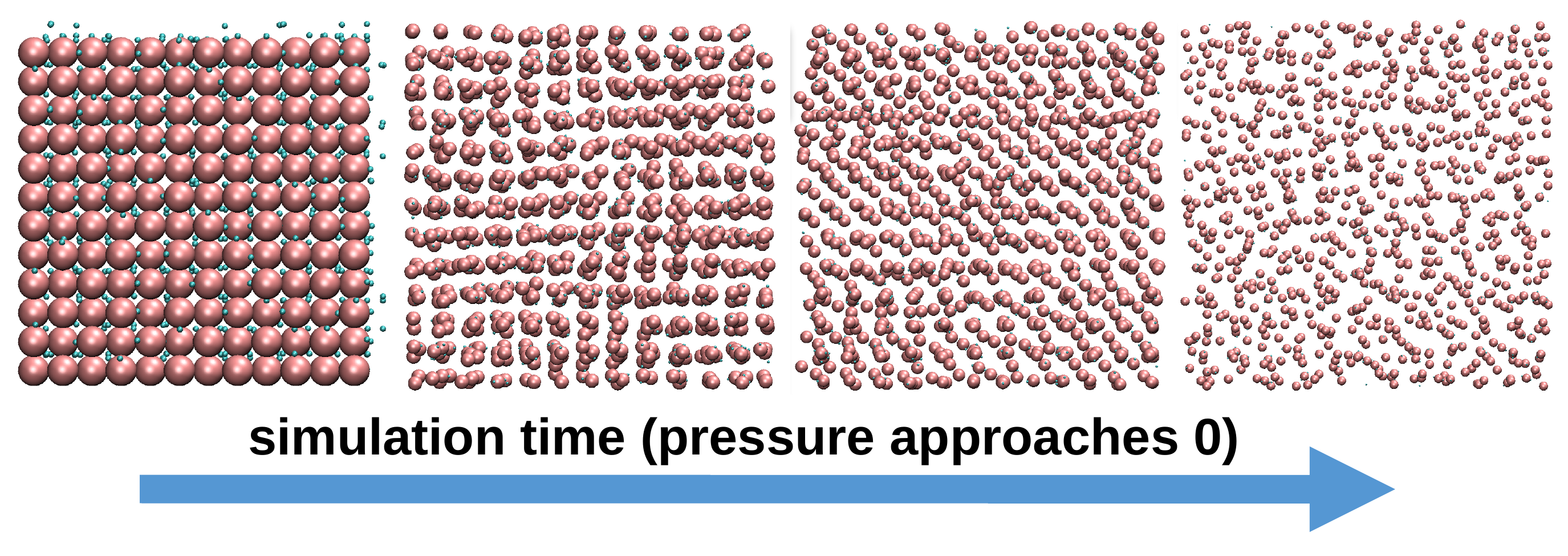}
\caption{An example of the melting process of unstable colloidal crystals as pressure approaches $0$ in NPT simulations. The simulation box is expanding to infinity simultaneously; here $\kappa\sigma=0.1$ and $q_B^*=+11$, $N_B/N_A=8$, $T^*=0.3$, and $q_A^*=-247$.}
\end{figure}

\subsection*{3. Diffusion Coefficient}
\begin{figure}[H]
\centering
\includegraphics[width=0.95\linewidth]{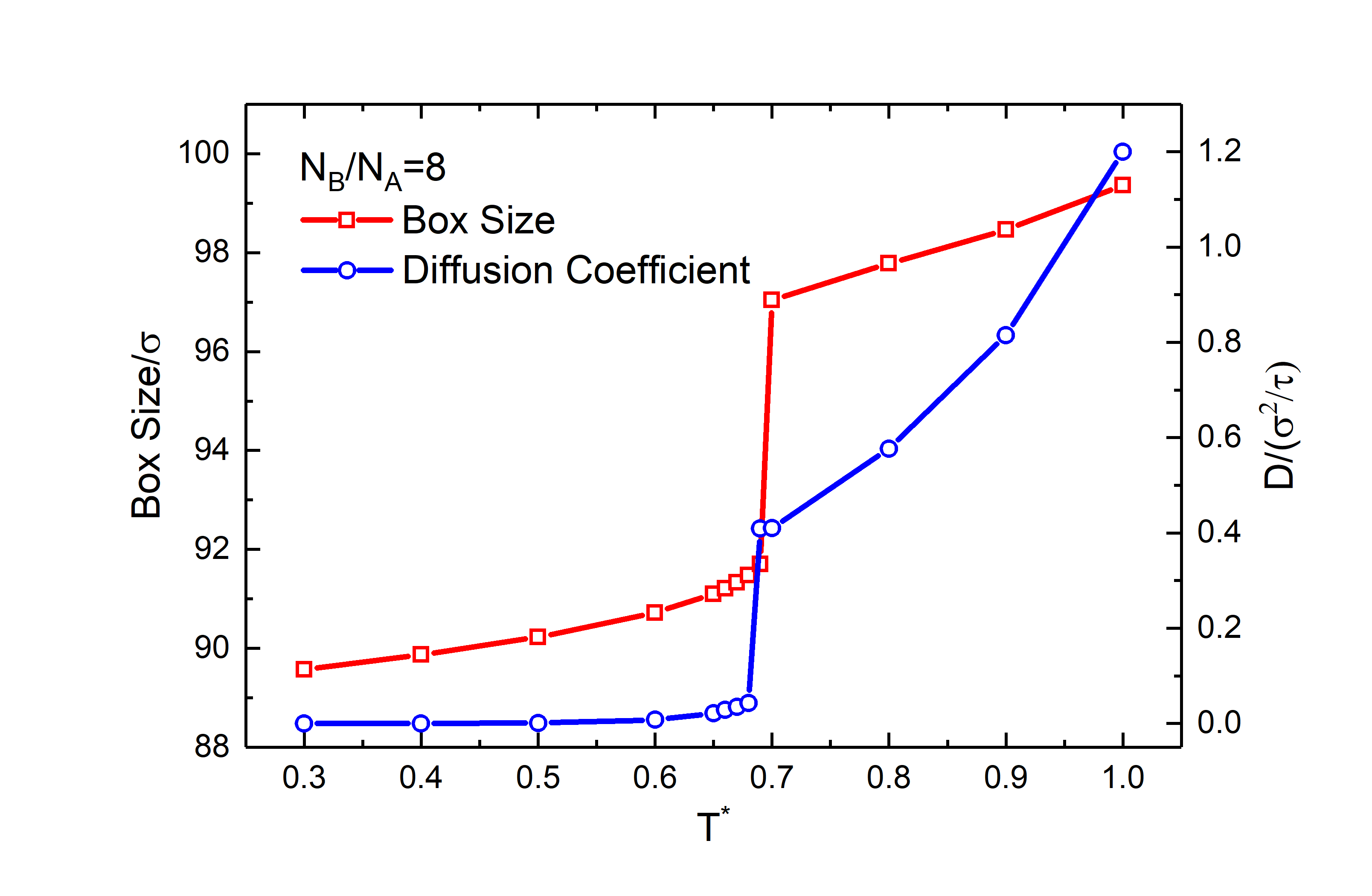}
\caption{Heating curves of the box size and the diffusion coefficient of small particles at $N_B/N_A = 8$. The diffusion coefficient also leaps at the ``ionic-superionic" transition temperature. The diffusion coefficient is calculated from the mean-square-displacement (MSD) of small particles $\langle r(t)^2\rangle$ over $5\times 10^6$ timesteps using $D=\langle r(t)^2\rangle/6t$.}
\end{figure}

\subsection*{4. Two Coexisting Ionic Phases}
\begin{figure}[H]
\centering
\includegraphics[width=0.95\linewidth]{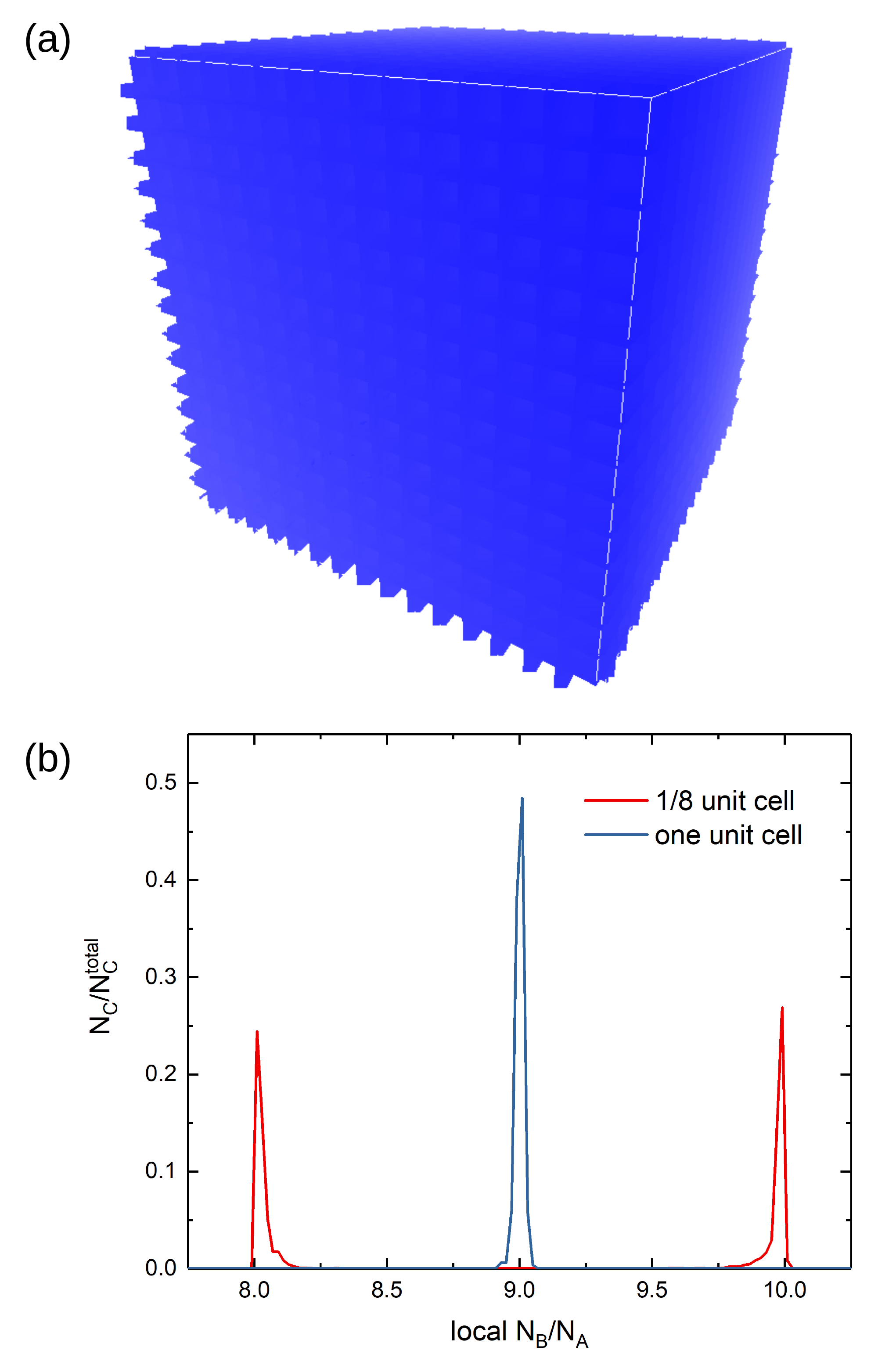} 
\caption{The $N_B/N_A=8$ and $N_B/N_A=10$ ionic phases both exist as microphases when they coexist (overall number ratio $N_B/N_A=9$, $T^*=0.2$). (a) Distribution of sub-unit cubes (1/8 unit cell) with local $N_B/N_A=8$ in the crystal. All the cubes satisfying $\left|N_B/N_A-8\right|<0.2$ are colored blue while the rest are left blank. Results show that these $N_B/N_A=8$ ionic cubes are nearly evenly dispersed within the crystal, hence the whole simulation box is colored blue. This is because at low temperatures the rearrangement of small particles can only happen between two neighboring cubes, i.e., two neighboring cubes with initial local number ratio 9/1 become one 8/1 cube and one 10/1. (b) The local number ratio calculation using two different cube sizes: 1/8 unit cell (red) and one unit cell (blue). The two separate peaks (red) merge into one single peak (blue) when using more coarse grained cubes, showing that these two ionic phases are both in micro\textendash size.}
\end{figure}

\subsection*{5. Justification of the parameters}
Here we justify that the parameters used in our simulations can be converted to reasonable experimental values, which is helpful for testing our results in experiments. There are many different ways to convert the quantities from the reduced units used in our simulations to the real units, and one possible way of conversions we provide is:
\begin{itemize}
  \item the distance unit: $\sigma$=1nm;
  \item the reduced temperature: $T^*=T/(428.6K)$ where $T$ is the real temperature and $K$ is Kelvin; the transition temperature in $N_B/N_A=8$ systems is about $T^*\approx0.7$, and we assume it corresponds to the room temperature, $300K$; hence $T^*/0.7=T/(300K)$ and then $T^*=T/(428.6K)$;
  \item the energy unit: $\varepsilon=k_B\times428.6K=5.9\times10^{-21}$ Joule; here $k_B=1.38\times10^{-23}Joule/K$ is the Boltzmann constant;
  \item the reduced charge: $q^*=q^r/(0.161\sqrt{\epsilon_r}e)$, where $q^r$ is the real charge, $e$ is the elementary charge, and $\epsilon_r$ is the dielectric constant of the media; this relationship is obtained by plugging the above quantities into $q^*=q/\sqrt{4\pi\epsilon_0\epsilon_r\sigma\varepsilon}$;
  \item the reduced pressure: $P^*=P\sigma^3/\varepsilon=P/(5.9\times10^6Pa)$ where $Pa$ is Pascal.
\end{itemize}
From these conversions, we have the particle sizes $R_A=5\sigma=5nm, R_B=1\sigma=1nm$, and the screening constant $\kappa=0.7nm^{-1}$ which corresponds to a $44mM$ NaCl salt solution. For the charges, assuming the media is water and $\epsilon_r=80$, plugging in $q_A^*=-247, q_B^*=+11$ we have:
\begin{equation}
 q_A^r=q_A^*(1+\kappa R_A)/e^{\kappa R_A}\times0.161\sqrt{\epsilon_r}e=-48e\nonumber
\end{equation}
\begin{equation}
 q_B^r=q_B^*(1+\kappa R_B)/e^{\kappa R_B}\times0.161\sqrt{\epsilon_r}e=+13e\nonumber
\end{equation}
Note that we are using the effective reduced charges, which need to be first converted to the bare charges via DLVO and then further converted to the real charges. In real units, the electrostatic interaction are governed by:
\begin{equation}
 U(r_{ij})=\frac{q_i^re^{\kappa R_i}q_j^re^{\kappa R_j}e^{-\kappa r_{ij}}}{4\pi\epsilon_0\epsilon_r(1+\kappa R_i)(1+\kappa R_j)r_{ij}}
\end{equation}

All the parameters, after converted in real units, are achievable in experiments. Therefore, in order to verify the ionic-superionic transition found in our simulations, experimentalists can prepare two kinds of particles with these given size and charge values, mix them in a 44mM NaCl salt solution, and the transition may be seen at around 300K.


\bibliographystyle{apsrev4-2}
\bibliography{SublatticeMelting}

\end{document}